\documentclass{article}
\usepackage{amsmath,amsfonts,amssymb,setspace,comment,url}
\usepackage{geometry}
\usepackage{cite}
\usepackage{xcolor}
\newcommand{\be}{\begin{equation}}
\newcommand{\ee}{\end{equation}}

\newcommand{\EG}[1]{\mathrm{E}_{#1(#1)}}

\newcommand{\SU}[1]{\mathrm{SU}( #1 )}
\newcommand{\SL}[1]{\mathrm{SL}( #1 )}

\newcommand{\SO}[1]{\mathrm{SO}( #1 )}

\newcommand{\Spin}[1]{\mathrm{Spin}(#1)}
\newcommand{\GH}{\mathrm{G}_{\mathrm{half}}}
\newcommand{\GR}{\mathrm{G}_{R}}

\newcommand{\USp}[1]{\mathrm{USp}(#1)}

\newcommand{\hK}{\hat{K}}
\newcommand{\K}{K}
\newcommand{\J}{J}
\newcommand{\hJ}{\hat{J}}

\newcommand{\bomega}{\bar{\omega}}
\newcommand{\homega}{\omega}

\newcommand{\hsigma}{\hat{\sigma}}
\newcommand{\comega}{\hat{\omega}}
\newcommand{\hphi}{\hat{\phi}}

\newcommand{\obf}[1]{\overline{\mathbf{#1}}}
\newcommand{\mbf}[1]{\mathbf{#1}}
\newcommand{\gL}{\mathcal{L}}

\newcommand{\gM}{\mathcal{M}}

\newcommand{\z}{z}

\newcommand{\bDelta}{\bar{\Delta}}
\numberwithin{equation}{section}

\newcommand\Tstrut{\rule{0pt}{3ex}}         
\newcommand\Bstrut{\rule[-1.3ex]{0pt}{0pt}}   

\begin{document}

\begin{titlepage}
\vfill

\begin{flushright}
	LMU-ASC 54/18\\
	MPP-2018-212
\end{flushright}

\vfill

\begin{center}
   \baselineskip=16pt
   	{\Large \bf Supersymmetric AdS$_{7}$ and AdS$_6$ vacua and their minimal consistent truncations from exceptional field theory}
   	\vskip 2cm
   	{\large \bf Emanuel Malek$^a$\footnote{\tt E.Malek@lmu.de}, Henning Samtleben$^b$\footnote{\tt Henning.Samtleben@ens-lyon.fr}, Valent\'{i} Vall Camell$^{a,c}$\footnote{\tt V.Vall@physik.uni-muenchen.de}}
   	\vskip .6cm
   	{\it $^a$ Arnold Sommerfeld Center for Theoretical Physics, Department f\"ur Physik, \\ Ludwig-Maximilians-Universit\"at M\"unchen, Theresienstra{\ss}e 37, 80333 M\"unchen, Germany \\ \ \\
    \it $^b$ Univ Lyon, Ens de Lyon, Univ Claude Bernard, CNRS,\\
    Laboratoire de Physique, F-69342 Lyon, France \\ \ \\
    \it $^c$ Max-Planck-Institut f\"ur Physik, Werner-Heisenberg-Institut, \\ F\"ohringer Ring 6, 80805 M\"unchen, Germany \\ \ \\}
   	\vskip 1cm
\end{center}
\vfill

\begin{abstract}
	We show how to construct supersymmetric warped AdS$_7$ vacua of massive IIA and AdS$_6$ vacua of IIB supergravity, using ``half-maximal structures'' of exceptional field theory. We use this formalism to obtain the minimal consistent truncations around these AdS vacua.
\end{abstract}
\vskip 4cm

\vfill

\end{titlepage}

\newpage

\section{Introduction}

Many supersymmetric AdS vacua of 10- and 11-dimensional SUGRA are known and play a crucial role in the AdS/CFT correspondence. Yet, we do not currently have a systematic geometric understanding of such vacua. Unlike fluxless supersymmetric Minkowski vacua, which are described by integrable $G$-structures and thus special holonomy manifolds, supersymmetric AdS vacua require non-vanishing fluxes and thus are not described by integrable $G$-structures in Riemannian geometry. However, recently a description of supersymmetric AdS vacua in terms of generalised $G$-structures has appeared using Exceptional Field Theory (ExFT) \cite{Berman:2010is,Berman:2011cg,Berman:2012vc,Hohm:2013pua}, focusing on the case of half-maximal supersymmetry \cite{Malek:2016bpu,Malek:2016vsh,Malek:2017njj}, and Exceptional Generalised Geometry (EGG) \cite{Coimbra:2011ky,Coimbra:2012af}, focusing on 1/4-maximal supersymmetry \cite{Ashmore:2015joa,Ashmore:2016qvs,Ashmore:2016oug} and more general analyses \cite{Coimbra:2014uxa,Coimbra:2015nha}\footnote{\cite{Coimbra:2017fqv} pointed out that there may be supersymmetric AdS vacua in even dimensions that are not described by generalised $G$-structures. However, these AdS vacua are unlikely to be complete and regular}. ExFT and EGG provide a unified description of metric and flux degrees of freedom of 10-/11-dimensional SUGRA by using enlarged geometric structures called generalised tangent bundles. It is on these enlarged bundles that the generalised $G$-structures can be defined.

This reformulation of SUGRA has already been succesfully used to study flux vacua. For example, it has led to maximally supersymmetric consistent truncations \cite{Aldazabal:2011nj,Geissbuhler:2011mx,Berman:2012uy} which uplift several interesting lower-dimensional gauged supergravities and their vacua to 10-/11-dimensional SUGRA \cite{Lee:2014mla,Hohm:2014qga,Malek:2015hma,Baguet:2015iou,Baguet:2015sma,Cassani:2016ncu,Inverso:2016eet,Inverso:2017lrz,Malek:2017cle}. Recently, \cite{Malek:2016bpu,Malek:2017njj} showed how to generalise this procedure to construct consistent truncations of 10-/11-dimensional SUGRA which break half the supersymmetry. Furthermore, \cite{Malek:2017njj} proved that for each warped half-maximally supersymmetric AdS$_D$ vacuum of 10-/11-dimensional SUGRA, there exists a consistent truncation to $D$-dimensional half-maximal gSUGRA containing only the graviton supermultiplet, thereby proving the half-maximal case of the conjecture \cite{Gauntlett:2007ma}. Such consistent truncations are particularly useful for the AdS/CFT correspondence where they allow us to study AdS vacua of 10-/11-dimensional SUGRA using lower-dimensional gauged SUGRAs. This can be used to study deformations of the AdS vacua, e.g. finding domain-wall solutions which are holographically dual to RG flow.

In this paper, we will show that generalised $G$-structures can be used to efficiently construct supersymmetric AdS vacua of 10-/11-dimensional SUGRA. We will focus on AdS$_7$ vacua of massive IIA and AdS$_6$ vacua of IIB SUGRA. A family of infinitely many such solutions have recently been found \cite{Apruzzi:2013yva,DHoker:2016ujz,DHoker:2017mds,DHoker:2017zwj}, parameterised by a cubic function in the case of AdS$_7$ and two holomorphic functions for AdS$_6$. We will show how these solutions can easily be constructed in terms of a universal half-maximal structure underlying them.

A further benefit of our approach is that, as shown in \cite{Malek:2017njj}, once we have described a supersymmetric AdS$_D$ vacuum by generalised $G$-structures, we immediately obtain a consistent truncation of the higher-dimensional SUGRA around the AdS vacuum to the minimal $D$-dimensional gauged SUGRA containing only the graviton supermultiplet. Thus, we will rederive the consistent truncation around the AdS$_7$ vacua of \cite{Passias:2015gya} and construct the minimal consistent truncation around the AdS$_6$ vacua. The fact that these vacua are described by a universal half-maximal structure explains the universality of their truncation Ansatz, which takes the same form for the entire family of solutions.

The outline of the paper is as follows. In section \ref{s:AdSReview}, we review the description of half-maximally supersymmetric AdS vacua using generalised $G$-structures \cite{Malek:2017njj}, and how to construct the minimal consistent truncations of these vacua. Next, we show how to calculate the generalised metric, which encode the SUGRA fields, from the half-maximal structures in section \ref{s:GenMetric}. We then show how to construct the family of infinitely-many AdS$_7$ vacua of mIIA and AdS$_6$ vacua of IIB using this method in sections \ref{s:AdS7} and \ref{s:AdS6}. Finally, in \ref{s:ConsTruncation}, we derive the minimal consistent truncation around these vacua before concluding in section \ref{s:Conclusions}.

\paragraph{Note added:} While finalising this manuscript, the paper \cite{Hong:2018amk} appeared which also constructs the minimal consistent truncation around supersymmetric AdS$_6$ vacua that we derive in section \ref{s:AdS6ConsTruncation}.

\section{Half-maximal AdS vacua from ExFT} \label{s:AdSReview}
Supersymmetric AdS vacua can be naturally described in ExFT using the language of generalised G-structures, analogous to the description of special holonomy spaces in Riemannian geometry.  In particular, as was shown in \cite{Malek:2016bpu,Malek:2016vsh,Malek:2017njj}, flux geometries of type II or 11-dimensional SUGRA admitting a half-maximal set of spinors can be described by a set of generalised tensors satisfying certain algebraic conditions. Here we will consider 10/11-dimensional geometries consisting of warped products $M_D \times M_{int}$, where $M_D$ denotes the external spacetime, $M_{int}$ is the internal space and we will focus on $D = 6, 7$ in this paper.

In order for $M_D$ to be a half-maximal AdS vacuum, $M_{int}$ must admit a set of nowhere-vanishing $d-1$ generalised vector fields $\J_u \in \Gamma\left({\cal R}_1\right)$, where $d = 11 - D$ and $u = 1, \ldots, d-1$, as well as a generalised tensor field $\hK \in \Gamma\left({\cal R}_{D-4}\right)$ satisfying
\begin{equation}
 \begin{split}
  \J_u \wedge \J_v - \frac1{d-1} \delta_{uv}\, \J_w \wedge \J^w &= 0 \,, \\
  \J_u \wedge \J^u \wedge \hK > 0 \,, \\
  \hK \otimes \hK \vert_{R_c} = 0 \,.
 \end{split} \label{eq:AlgConditions}
\end{equation}
Here ${\cal R}_i$ are different generalised tangent bundles whose fibre is the $R_i$ representation of $\EG{d}$, listed in table \ref{t:Reps}, where also the representation $R_c$ is given. The $\wedge$ products map
\begin{equation}
 \begin{split}
  \wedge: {\cal R}_i &\otimes {\cal R}_j \longrightarrow {\cal R}_{i+j} \quad \textrm{ when } i + j < D - 2 \,, \\
  \wedge: {\cal R}_i &\otimes {\cal R}_{D-2-i} \longrightarrow {\cal S} \,,
 \end{split}
\end{equation}
where ${\cal S}$ denotes the space of scalar densities of weight 1 under generalised diffeomorphisms. The explicit expressions of $\wedge$ in terms of $\EG{d}$ invariant is given in \cite{Malek:2017njj}.

\vskip1em
\begin{table}[h]\centering
	\begin{tabular}{|c|c|c|c|c|c||c|}
		\hline
		$D$ & $\EG{d}$ & $R_1$ & $R_2$ & $R_3$ & $R_4$ & $R_c$ \Tstrut\Bstrut \\ \hline
		7 & $\SL{5}$ & $\mbf{10}$ & $\obf{5}$ & $\mbf{5}$ & $\obf{10}$ & $\emptyset$ \\
		6 & $\Spin{5,5}$ & $\mbf{16}$ & $\mbf{10}$ & $\obf{16}$ & N/R & $\mbf{1}$ \\
		\hline
	\end{tabular} \label{t:Reps}
	\vskip-0.5em
	\caption{\small{Different representations of the exceptional groups that are relevant to half-maximal structures.}}
\end{table}

Throughout, we will raise and lower the $u, v = 1 \ldots, d - 1$ indices using $\delta_{uv}$. A set of generalised tensors $\left(\J_u, \hK\right)$ as above are called a half-maximal structure and are stabilised by a $\GH = \SO{d-1}$ subgroup of the exceptional group $E_{d(d)}$ given in table \ref{t:Edd}. The maximal commutant of $\SO{d-1} \subset E_{d(d)}$ is $\SO{d-1}_R$ and rotates the $d-1$ $J_u$'s amongst each other.

From the half-maximal structure one can also define the following generalised tensors that will be useful in the following:
\begin{equation}
\J_u \wedge \J_v = \delta_{uv} \K \,, \qquad \K \wedge \hK = \kappa^{D-2} \,, \qquad \hJ_u = J_u \wedge \hK \,, \label{eq:Kkappa}
\end{equation}
where $\K \in \Gamma\left({\cal R}_2\right)$, $\kappa$ is a scalar density of weight $\frac{1}{D-2}$ and $\hJ_u \in \Gamma\left({\cal R}_{D-3}\right)$. The explicit expressions for the above equations \eqref{eq:AlgConditions} and \eqref{eq:Kkappa} in terms of $\EG{d}$-invariants can be found in \cite{Malek:2017njj}.

Furthermore, the BPS equations for the AdS$_D$ vacuum are equivalent to the following differential equations
\begin{equation}
 \begin{split}
  \gL_{\J_u} \J_v &= - \Lambda_{uvw} J^w \,, \\ 
  \gL_{\J_u} \hK &= 0 \,, \\
  d\hK &= \left\{ \begin{array}{c}
  \frac{1}{3!\sqrt{2}} \epsilon^{uvw} \Lambda_{uvw} \K \,, \quad \textrm{when D = 7} \,, \\
  \frac{1}{9} \epsilon_{uvwx} \Lambda^{uvw} \J^x \,, \quad \, \textrm{ when D = 6} \,,
  \end{array} \right.
 \end{split} \label{eq:DiffConditions}
\end{equation}
where $\Lambda_{uvw} = \Lambda_{[uvw]}$ are completely antisymmetric, $\gL$ denotes the generalised Lie derivative of ExFT \cite{Berman:2011cg,Coimbra:2011ky,Berman:2012vc} and $d: \Gamma\left({\cal R}_{D-4}\right) \longrightarrow \Gamma\left({\cal R}_{D-5}\right)$ is a certain nilpotent operator as defined in \cite{Malek:2016bpu,Malek:2017njj} and which also appears in studies of the tensor hierarchy of ExFT \cite{Cederwall:2013naa,Hohm:2015xna,Wang:2015hca}.

\vskip1em
\begin{table}[h]\centering
			\begin{tabular}{|c|c|c|c|c|}
				\hline
				$D$ & $\EG{d}$ & $H_d$ & $\GH$ & $\GR$ \Tstrut\Bstrut \\ \hline
				7 & $\SL{5}$ & $\USp{4}$ & $\SU{2}$ & $\SU{2}$ \\
				6 & $\Spin{5,5}$ & $\USp{4}\times\USp{4}$ & $\SU{2}\times\SU{2}$ & $\SU{2}\times\SU{2}$ \\
				\hline
			\end{tabular} \label{t:Edd}
			\vskip-0.5em
			\caption{\small{$\GH$ structures and R-symmetry groups in $6$ and $7$ dimensions. In six dimensions, one can also have chiral half-maximal supersymmetry but we ignore this here since there are no chiral AdS$_6$ vacua. The interested reader is referred to \cite{Malek:2017njj} for more information.}}
\end{table}

The differential conditions \eqref{eq:DiffConditions} encode the BPS conditions for the supersymmetric AdS$_D$ vacuum in a geometric language. As we will see the algebraic and differential conditions \eqref{eq:AlgConditions} and \eqref{eq:DiffConditions} can easily be solved in a variety of different cases, providing an efficient way of constructing supersymmetric AdS vacua of 10/11-dimensional supergravity. Moreover, once we have the half-maximal structure for an AdS vacuum, we can immediately construct a consistent truncations around the AdS vacuum to the $D$-dimensional half-maximal gauged SUGRA containing only the graviton supermultiplet \cite{Malek:2016bpu,Malek:2017njj}, as we review in section \ref{s:RevConsTruncation}. We will use this method to find the minimal consistent truncations around supersymmetric AdS$_6$ vacua of IIB SUGRA, as well as derive the consistent truncations around supersymmetric AdS$_7$ vacua of massive IIA SUGRA, where our expressions agree with \cite{Passias:2015gya}.

The half-maximal structure encodes the 11-dimensional / type II supergravity fields, just like a complex and K\"ahler structure encode the metric. In ExFT, the supergravity fields parameterise the generalised metric $\gM_{MN}$ which lives in the coset space
\begin{equation}
 \gM_{MN} \in \frac{\EG{d}}{H_d} \,,
\end{equation}
where $H_d$ is the maximal compact subgroup of $\EG{d}$\footnote{Here we will be careless about discrete factors and not differentiate between $H_d$ and its double cover}, listed in table \ref{t:Edd}. As we will show in section \ref{s:GenMetric}, the generalised metric $\gM_{MN}$ can be expressed in terms of an $\SO{d-1}_R$-invariant combination of the half-maximal structure $\J_u$ and $\hK$, just like it can be expressed in terms of generalised vielbeine $\gM_{MN} = E_M{}^{\bar{M}} E_{N}{}^{\bar{N}} \delta_{\bar{M}\bar{N}}$.

Before we proceed to discuss the AdS$_{6,7}$ cases in detail, we will first make some general observations about the differential conditions \eqref{eq:DiffConditions} and what they imply for $\J_u$ and $\hK$. We see that the generalised Lie derivative of the $\J_u$'s generates an $\SO{d-1}_R$ rotation under which the $\J_u$'s transform in the vector representation while $\hK$ is invariant. However, as mentioned above, the generalised metric and hence the SUGRA fields are constructed from $\SO{d-1}_R$-invariant combinations of the $\J_u$'s and $\hK$. Therefore,
\begin{equation}
 \gL_{\J_u} \gM_{MN} = 0 \,,
\end{equation}
and the $\J_u$ are generalised Killing vector fields. Generalised vector fields are a formal sum of spacetime vector fields plus certain differential forms. Therefore, for $\J_u$ to be generalised Killing, implies that either they consist of non-zero spacetime Killing vector fields with accompanying gauge transformations such that the gauge potentials are left invariant, or they have an identically vanishing spacetime vector field part and consist of trivial gauge transformations, i.e. exact differential forms. For such a ``trivial'' generalised Killing vector field $V$ we would have
\begin{equation}
 \gL_{V} = 0 \,,
\end{equation}
acting on any tensor. We will make use of this general insight in sections \ref{s:AdS7} and \ref{s:AdS6} when constructing the AdS vacua.

\subsection{Minimal consistent truncation} \label{s:RevConsTruncation}
Once we have constructed the half-maximal structure $\J_u$ and $\hK$ corresponding to a half-maximal AdS$_D$ vacuum, we can immediately construct a consistent truncation around this vacuum to a minimal half-maximal $D$-dimensional SUGRA \cite{Malek:2017njj}. That such a consistent truncation should always exist for any warped supersymmetric AdS vacuum of 10-/11-dimensional SUGRA was conjectured in \cite{Gauntlett:2007ma} and proven in the half-maximal case for $D \geq 4$ in \cite{Malek:2017njj}.

The truncation Ansatz is linear on the half-maximal structure and given as follows. We denote by $Y^M$ the internal coordinates on $M_{int}$ and by $x^\mu$ the external coordinates on $M_D$. Then, the truncation Ansatz (of the scalar sector) is given by \cite{Malek:2016bpu,Malek:2017njj}
\begin{equation}
 \begin{split}
  {\cal J}_u(x,Y) &= X^{-1}(x)\, J_u(Y) \,, \\
  {\cal \hK}(x,Y) &= X^2(x)\, \hK(Y) \,,
 \end{split} \label{eq:TruncAnsatz}
\end{equation}
where $X(x)$ is the scalar field of the $D$-dimensional half-maximal SUGRA. The consistency of the truncation Ansatz is guaranteed by the differential conditions \eqref{eq:DiffConditions} satisfied by the $J_u$, $\hK$ as shown in \cite{Malek:2017njj}. Upon truncation, $X(x)$ becomes the scalar field of the minimal half-maximal $D$-dimensional gauged SUGRA with embedding tensor given by $\Lambda_{uvw}$ in \eqref{eq:DiffConditions}.

The consistent truncation can easily be extended to the other fields of the $D$-dimensional SUGRA as explained in \cite{Malek:2017njj}. However, since non-vanishing vacuum expectation values of these fields will typically break Lorentz symmetry, we will not include them in this paper.

\section{Generalised metric from the half-maximal structure} \label{s:GenMetric}
As we mentioned above, the half-maximal structure determines the supergravity fields which in ExFT are encoded in the generalised metric. We therefore need to find a way to compute the generalised metric from the half-maximal structure $\J_u$, $\hK$. The generalised metric parameterises the coset space
\begin{equation}
\gM_{MN} \in \frac{\EG{d}}{H_d} \,,
\end{equation}
and hence must be an $\EG{d}$ group element that is invariant under $H_d$. Since $\J_u$ and $\hK$ are by construction invariant under $\GH = \SO{d-1} \subset H_d$, we must construct $\gM_{MN}$ using an $\SO{d-1}_R$-invariant combination of $\J_u$ and $\hK$.

\subsection{Generalised metric in $\SL{5}$ ExFT}
In $\SL{5}$ ExFT \cite{Berman:2010is,Berman:2011cg,Musaev:2015ces}, the generalised metric is often used either in the $R_1 = \mbf{10}$ representation or in the fundamental representation, $R_2 = \mbf{5}$, of $\SL{5}$. The two are related by
\begin{equation}
 \gM_{ab,cd} = 2 \gM_{a[c}\gM_{d]b} \,,
\end{equation}
where $a, b = 1, \ldots, 5$ denote fundamental $\SL{5}$ indices \cite{Berman:2011cg}.

The generalised metric and its inverse are given by
\begin{equation}
 \begin{split}
  \gM_{ab,cd} &= 8 \, \kappa^{-8} \hJ_{u\,ab} \hJ^{u}{}_{cd} - \kappa^{-3} \epsilon_{abcde} \hK^e - \frac{1}{6\sqrt{2}} \kappa^{-3} \epsilon^{uvw} \epsilon_{abefg} \epsilon_{cdhij} \J_{u}{}^{ef} \J_{v}{}^{hi} \J_w{}^{gj} \,, \\
  \gM^{ab,cd} &= 2 \, \kappa^{-2} \J_{u}{}^{ab} \J^{u,cd} - \kappa^{-2} \epsilon^{abcde} \K_e - \frac{2\sqrt{2}}{3} \kappa^{-12} \epsilon^{uvw} \epsilon^{abefg} \epsilon^{cdhij} \hJ_{u\,ef} \hJ_{v\,hi} \hJ_{w\,gj} \,,
 \end{split} \label{eq:SL510GM}
\end{equation}
where $\hJ_{u\,ab}$ and $\kappa$ are defined as in \eqref{eq:Kkappa} which here become
\begin{equation}
\hJ_{u\,ab} = \frac14 \epsilon_{abcde} \J_u{}^{cd} \hK^e \,, \qquad \kappa^5 = \frac{1}{12} \epsilon_{abcde} \J_u{}^{ab} \J^{u}{}^{cd} \hK^{e} \,,
\end{equation}
where $\epsilon_{abcde}$ is the constant $\SL{5}$-invariant tensor.
Similarly, the generalised metric and its inverse in the $\mbf{5}$ representation of $\SL{5}$ are given by
\begin{equation}
 \begin{split}
  \gM_{ab}& = \kappa^{-4} \left( \K_a \K_b + \frac{4\sqrt{2}}{3} \kappa^{-5} \, \epsilon^{uvw} \hJ_{u,ac} \hJ_{v,bd} \J_w{}^{cd} \right) \,, \\
  \gM^{ab}& = \kappa^{-6} \left( \hat{\K}^a \hat{\K}^b+  \frac{2\sqrt{2}}{3} \, \epsilon^{uvw}  \J_u{}^{ac} \J_v{}^{bd} \hJ_{w,cd} \right) \,.
 \end{split} \label{eq:SL55GM}
\end{equation}

\subsection{Generalised metric in $\SO{5,5}$ ExFT}
In $\SO{5,5}$ ExFT \cite{Berman:2011pe,Abzalov:2015ega}, the generalised metric is often used either in the $R_1 = \mbf{16}$ representation or in the fundamental representation, $R_2 = \mbf{10}$, of $\SO{5,5}$. The two are related by
\begin{equation}
 \gM_{MP} \gM_{NQ} \left(\gamma_I\right)^{MN} \gM^{IJ} = \left(\gamma^J\right)_{PQ} \,,
\end{equation}
where $M = 1, \ldots, 16$ label the $\mbf{16}$ representation of $\SO{5,5}$, $I = 1, \ldots, 10$ labels the $\mbf{10}$ representation of $\SO{5,5}$ and $\left(\gamma_I\right)^{MN}$ and $\left(\gamma_I\right)_{MN}$ are the $\SO{5,5}$ $\gamma$-matrices satisfying
\begin{equation}
 \left( \gamma_I\right)^{MP} \left( \gamma_J \right)_{NP} + \left( \gamma_J \right)^{MP} \left( \gamma_I \right)_{NP} = 2\, \eta_{IJ} \delta^{M}_N \,,
\end{equation}
where $\eta_{IJ}$ is the constant $\SO{5,5}$-invariant matrix with which we raise/lower fundamental $\SO{5,5}$ indices. Furthermore, $\gM_{IJ}$ must satisfy
\begin{equation}
 \gM_{IK} \gM_{JL} \eta^{KL} = \eta_{IJ} \,.
\end{equation}

We thus find the generalised metric and its inverse in the $\mbf{16}$ are given by
\begin{equation}
 \begin{split}
  \gM_{MN} &= \frac{1}{\sqrt{2}} \left(4\, \kappa^{-6}\, \hJ^u{}_M \hJ_{u\,N} - \kappa^{-2} \left(\gamma^I\right)_{MN} \hK_I \right. \\
  & \quad \left. -\frac1{4!}\, \kappa^{-6} \epsilon^{uvwx} \left(\gamma_I\right)_{MP} \left(\gamma_J\right)_{NQ} \left(\gamma^{IJ}\right)^{S}{}_R \J_u{}^P \J_v{}^Q \J_w{}^R \hJ_{x,S} \right) \,, \\
  \gM^{MN} &= \frac{1}{\sqrt{2}} \left( 2\, \kappa^{-2} \J_u{}^M \J^{u\,N} - \kappa^{-2} \left(\gamma_I\right)^{MN} \K^I \right. \\
  & \left. \quad - \frac{2}{4!} \kappa^{-10} \epsilon_{uvwx} \left(\gamma_I\right)^{MP} \left(\gamma_J\right)^{NQ} \left(\gamma^{IJ}\right)^{S}{}_R \hJ^u{}_P \hJ^v{}_Q \J_w{}^R \hJ_{x,S} \right) \,,
 \end{split} \label{eq:SO55GenMetric1}
\end{equation}
where $\hJ_{u\,M}$ and $\kappa$ are defined in \eqref{eq:Kkappa}, and here given explicitly by
\begin{equation}
 \hJ^u{}_M =\frac12 \left(\gamma^I\right)_{MN} \hK_I \J^{u\,N} \,, \qquad \kappa^4 = \frac18 \left(\gamma^I\right)_{MN} \J_u{}^{M} \J^{u}{}^{N} \hK_I \,.
\end{equation}
Similarly, the generalised metric in the $\mbf{10}$ is
\begin{equation}
 \gM_{IJ} = \left( \frac{1}{4!} \epsilon^{uvwx} \left(\gamma_{IK} \right)_{M}{}^{N} \left( \gamma_J{}^K \right)_{P}{}^{Q} J_u{}^M \hJ_{v,N} \J_w{}^P \hJ_{x,Q} + \kappa^{-4}\, \K_I \K_J + \kappa^{-4} \hK_I \hK_J \right) \,. \label{eq:SO55GenMetric2}
\end{equation}

\section{AdS$_7$ vacua from massive IIA supergravity} \label{s:AdS7}
We will now show how to use this method to construct AdS$_7$ vacua of massive IIA SUGRA. First, we let $\Lambda_{uvw} = \sqrt{2} R^{-1} \epsilon_{uvw}$ so that the differential conditions \eqref{eq:DiffConditions} become
\begin{equation}
 \begin{split}
  \gL_{\J_u} \J_v &= -\frac{\sqrt{2}}{R}\, \epsilon_{uvw} \J^w\,, \\
  \gL_{\J_u} \hK &= 0 \,, \\
  d\hK &= \frac{1}{R}\, K \,.
 \end{split} \label{eq:SL5DC}
\end{equation}
We see that the $\J_u$'s generate $\SU{2}_R$ rotations via the generalised Lie derivative. $\J_u$ form triplets of $\SU{2}_R$, while $\hK$ are invariant. As we discussed in section \ref{s:AdSReview} this implies that $\J_u$ are generalised Killing vector fields. Since $\gL_{\J_u} \neq 0$, none of the $\J_u$ are trivial generalised Killing vector fields and hence must contain spacetime Killing vectors. From \eqref{eq:SL5DC} we see that these spacetime Killing vectors must generate an $\SU{2}_R$ algebra and hence are related to an $S^2$ geometry. Therefore, we will consider the internal space
\begin{equation}
 M_{int} = S^2 \times I \,,
\end{equation}
where $I$ is an interval with coordinate $z$, where in principle we allow off-diagonal metrics between the $S^2 \times I$ (although we will see that supersymmetry does not allow these off-diagonal terms). We will parameterise $S^2$ by the three functions $y_u$, $u = 1, \ldots, 3$ satisfying $y_u\, y^u = 1$. Further details of our $S^2$ convention can be found in appendix \ref{A:S2}.

\subsection{$\SL{5}$ ExFT and IIA SUGRA}
Supersymmetric AdS$_7$ vacua are characterised by three generalised vector fields $\J_u \in \Gamma\left({\cal R}_1\right)$ and a generalised tensor field $\hK \in \Gamma\left({\cal R}_3\right)$. In IIA SUGRA these become formal sums of spacetime vector fields and differential forms,
\begin{equation}
 \begin{split}
  \J_u &= V_u + \lambda_u + \sigma_u + \phi_u \,, \\
  \hK &= \homega_{(0)} + \homega_{(2)} + \homega_{(3)} \,,
 \end{split}
\end{equation}
where $V_u$, $\lambda_u$, $\sigma_u$ and $\phi_u$ are the vector, 1-form, 2-form and scalar parts of $\J_u$, while $\homega_{(p)}$ are the $p$-forms appearing in $\hK$. For completeness' sake, a generalised tensor $\K \in \Gamma\left({\cal R}_2\right)$ becomes
\begin{equation}
 \K = \bomega_{(0)} + \bomega_{(1)} + \bomega_{(3)} \,,
\end{equation}
where $\bomega_{(p)}$ are $p$-forms.

In IIA SUGRA, the wedge products appearing in the algebraic conditions \eqref{eq:AlgConditions} become
\begin{equation}
 \begin{split}
  \J_u \wedge \J_v &= 2 \imath_{V_{(u}} \lambda_{v)} - 2 \left( \lambda_{(u} \phi_{v)} + \imath_{V_{(u}} \sigma_{v)} \right) - 2 \lambda_{(u} \wedge \sigma_{v)} \,, \\
  \hK \wedge \K &= \homega_{(0)}\, \bomega_{(3)} + \homega_{(1)} \wedge \bomega_{(2)} + \bomega_{(0)}\, \homega_{(3)} \,.
 \end{split} \label{eq:mIIAAlgConditions}
\end{equation}
The quadratic algebraic constraint on $\hK$ is automatically fulfilled for $\SL{5}$ ExFT \cite{Malek:2017njj}. The differential operators appearing in the differential conditions \eqref{eq:SL5DC} become
\begin{equation}
 \begin{split}
  \gL_{J_u} J_v &= L_{v_u} V_v + L_{V_u} \lambda_v + L_{V_u} \sigma_v + L_{V_u} \phi_v \\
  & \quad + \imath_{V_v} \left( m \lambda_u - d\phi_u \right) - \imath_{V_v} \left( d\lambda_u \right) - \imath_{V_v} \left( d\sigma_u \right) + \phi_v \left( d\lambda_u \right) + \lambda_v \wedge \left( m \lambda_u - d\phi_u \right) \,, \\
  \gL_{J_u} \hK &= L_{V_u} \homega_{(0)} + L_{V_u} \homega_{(2)} + L_{V_u} \omega_{(3)} \\
  & \quad - \homega_{(0)} \left( d\lambda_u \right) - \homega_{(0)} \left( d\sigma_u \right) - \homega_{(2)} \wedge \left( m \lambda_u - d\phi_u \right) \,, \\
  d\hat{K} &= - d\homega_{(0)} + d \homega_{(2)} \,,
 \end{split} \label{eq:mIIADiffConditions}
\end{equation}
where we have included the Roman's mass $m$ as in \cite{Ciceri:2016dmd,Cassani:2016ncu}.

\subsection{Half-maximal structure}
Before continuing, we need to discuss the possible gauge potentials living in $M_{int}$. In IIA SUGRA, we need to consider a 2-form and 3-form field strength in $M_{int}$ which must form $\SU{2}_R$-symmetry singlets. The 2-form gauge potential can always be chosen to be an $\SU{2}_R$-symmetry singlet. However, the 1-form gauge potential $A$ will necessarily violate the $\SU{2}_R$-symmetry. As we will use R-symmetry as a guiding principle, we will have to include the 1-form gauge potential $A$ by hand as a ``twist term'', as e.g. in \cite{Lee:2014mla}. This implies that we take $\phi_u = \hphi_u + \imath_{V_u} A$ and $\sigma_u = \hsigma_u + \lambda_u \wedge A$ and $\homega_{(3)} = \comega_{(3)} + \homega_{(2)} \wedge A$. On the other hand, our Ansatz below will naturally incorporate the 2-form potential.

The most general $\J_u$ we can construct that is compatible with the $\SU{2}_R$ symmetry and that satisfies the algbraic conditions \eqref{eq:AlgConditions} is, up to generalised diffeomorphisms (i.e. gauge transformations and diffeomorphisms), given by
\begin{equation}
 \begin{split}
  \J_u &= \frac{2\sqrt{2}}{R} v_u + \frac{R}{4} \left( g(z)\, dy_u -\frac{h(z)}{q(z)} y_u\, dz \right) - \frac{R}{2} q(z)\,y_u + \frac{R^3}{16\sqrt{2}} \left( q(z)\,g(z)\,y_u\,vol_{S^2} + h(z)\, \theta_u \wedge dz \right) \\
  & \quad + \frac{2\sqrt{2}}{R} \imath_{v_u} A + \frac{R}{4} \left(g(z)\,dy_u - \frac{h(z)}{q(z)} y_u\, dz \right) \wedge A \,,
 \end{split} \label{eq:SL5Structure}
\end{equation}
where $v_u$ are Killing vectors and $\theta_u$ certain 1-forms on $S^2$ (see appendix \ref{A:S2}), and $g(z)$, $q(z)$, $h(z)$ are so far arbitrary functions of $z$. Furthermore, the most general $\hK$ constructed from R-symmetry singlets is, up to generalised diffeomorphisms, given by
\begin{equation}
 \begin{split}
  \hK &= \frac{R}{2}\,s(z) + \frac{R^3}{16\sqrt{2}} \left( g(z)\,s(z) - t(z) \right) vol_{S^2} + \frac{R^3}{16\sqrt{2}} \left(g(z)\,s(z) - t(z) \right) vol_{S^2} \wedge A \,.
 \end{split}
\end{equation}
The algebraic condition $\J_u \wedge \J^u \wedge \hK > 0$ now becomes
\begin{equation}
 \frac{R^5}{64\sqrt{2}}\, h(z)\,t(z)\, vol_{S^2} \wedge dz > 0 \,.
\end{equation}
Allowing for the $S^2$ to shrink at the boundary of the interval parameterised by $z$, we have
\begin{equation}
 h(z)\, t(z) \geq 0 \,,
\end{equation}
with equality at the boundary of $I$. Finally, as discussed $R$-symmetry implies that
\begin{equation}
 dA = R^2 l(z)\, vol_2 \,,
\end{equation}
for some $l(z)$. With \eqref{eq:SL5Structure}, the differential conditions \eqref{eq:mIIADiffConditions} reduce to
\begin{equation}
 \begin{split}
  m\, \lambda_u + \imath_{V_u} dA - d \hphi_u &= d\hsigma_u - \lambda_u \wedge dA = d\lambda_u = 0 \,, \\
  d\homega_{(0)} &= - \frac{2}{3R} \lambda_u \hphi^u\,, \\
  d\homega_{(2)} &= - \frac{2}{3R} \left( \imath_{V_u} \hsigma^u + \lambda_u \wedge \hsigma^u \right) \,.
 \end{split} \label{eq:mIIADC}
\end{equation}

We can always redefine the coordinate $z$ to make $h(z)$ any functions we choose. A particular convenient choice is to take $h(z) = q(z)$, whereupon the differential conditions \eqref{eq:mIIADC} become
\begin{equation}
 \dot{g} = -1 \,, \qquad \dot{q} = \frac{m}{2} \,, \qquad \dot{s} = q \,, \qquad \dot{t} = -s \,, \qquad l(z) = - \frac{q}{4\sqrt{2}} - \frac{m\,g}{8\sqrt{2}} \,.
\end{equation}
Without loss of generality we can integrate $\dot{g}= -1$ to $g = - z$, absorbing any constant of integration by shifting $z$. Furthermore, we can express $s$ and $q$ in terms of derivatives of $t$ which must satisfy
\begin{equation}
 \dddot{t} = - \frac{m}{2} \,,
\end{equation}
and $t \geq 0$ with equality at the boundary of $I$.

Altogether the half-maximal structure then becomes
\begin{equation}
 \begin{split}
  \J_u &= \frac{2\sqrt{2}}{R} v_u - \frac{R}{4} d\left(y_u\,z\right) + \frac{R}{2} \ddot{t}\,y_u + \frac{R^3}{16\sqrt{2}} \ddot{t} \left( d\left(z\, \theta_u \right) - z\,y_u\,vol_2 \right) \\
  & \quad + \frac{2\sqrt{2}}{R} \imath_{v_u} A - \frac{R}{4} d\left(y_u\,z\right) \wedge A \,, \\
  \hK &= -\frac{R}{2} \dot{t} + \frac{R^3}{16\sqrt{2}} \left( z\, \dot{t} - t \right) vol_{S^2} + \frac{R^3}{16\sqrt{2}} \left(z\,\dot{t} - t \right) vol_{S^2} \wedge A \,, \label{eq:AdS7Structures}
 \end{split}
\end{equation}
determined entirely by $t(z)$ satisfying
\begin{equation}
 \dddot{t} = -\frac{m}{2} \,, \qquad t \geq 0 \textrm{ with equality at } \partial I \,, \label{eq:tcond}
\end{equation}
and where
\begin{equation}
 dA = \frac{R^2}{4\sqrt{2}} \left( \ddot{t} + \frac{m}{2}\,z \right) vol_{S^2} \,.
\end{equation}

\subsection{The AdS$_7$ vacua}
The SUGRA fields with legs on $M_{int} = S^2 \times I$ can be read off from the generalised metric constructed from $\J_u$ and $\hK$ in \eqref{eq:AdS7Structures}. For this we use the parameterisation of generalised metric by IIA SUGRA fields given in \cite{Malek:2015hma}. The warp factor the AdS$_7$ metric is given by \cite{Malek:2016bpu,Malek:2017njj}
\begin{equation}
 w_7 = |g|^{1/5} \kappa^{-2} e^{-4\psi/5} \,,
\end{equation}
where $|g|$ is the determinant of the internal space in string frame and $\psi$ is the IIA dilaton. Thus, we find the infinite family of supersymmetric AdS$_7$ vacua determined by the function $t(z)$ satisfying \eqref{eq:tcond}.
\begin{equation}
 \begin{split}
  ds_{10}^2 &= R^2\,\sqrt{- \frac{t}{\ddot{t}}}\, ds_{AdS_7}^2 + \frac{R^2}{8} \sqrt{- \frac{\ddot{t}}{t}} \left( \frac{t^2}{\dot{t}^2 - 2\, \ddot{t}\, t} ds_{S^2}^2 + d\z^2 \right) \,, \\
  e^{\psi} &= \frac{2}{R} \left(- \frac{t}{\ddot{t}}\right)^{3/4} \frac{1}{\sqrt{\dot{t}^2 - 2\, \ddot{t} \, t}} \,, \\
  B_2 &= \frac{R^2}{8\sqrt{2}} \left( \z - \frac{\dot{t}\, t}{ \dot{t}^2 - 2 \, \ddot{t} \, t} \right) vol_{2} \,, \\
  F_2 &= \frac{R^2}{8\sqrt{2}} \left( 2 \ddot{t} + \frac{m\, \dot{t}\, t}{ \dot{t}^2 - 2 \, \ddot{t} \, t} \right) vol_{2} \,,
 \end{split}
\end{equation}
where the metric is expressed in string frame and $F_2 = dA - m\,B_2$ is the Ramond-Ramond 2-form field strength of mIIA SUGRA. This is clearly the family of AdS$_7$ solutions found in \cite{Apruzzi:2013yva} in the coordinate choice of \cite{Cremonesi:2015bld}, where our variables are related to theirs by the rescaling
\begin{equation}
 t = \frac{4\sqrt{2}}{81} \alpha \,, \qquad z = 2\sqrt{2}\pi Z\,.
\end{equation}

\section{AdS$_6$ vacua from IIB supergravity} \label{s:AdS6}
We next consider supersymmetric AdS$_6$ in IIB SUGRA. We begin by rewriting the differential conditions \eqref{eq:DiffConditions} by introducing the $\SO{4}_R$ vector $\Lambda_u$ defined as
\begin{equation}
\Lambda_{uvw} = \frac{3}{2^{1/4}} \epsilon_{uvwx} \Lambda^x \,.
\end{equation}
Then the differential conditions \eqref{eq:DiffConditions} become
\begin{equation}
 \begin{split}
  \gL_{\J_u} \J_v &= -\frac{3}{2^{1/4}} \epsilon_{uvwx} \J^w \Lambda^x \,, \\
  \gL_{\J_u} \hK &= 0 \,, \\
  d\hK &= 2^{3/4} \Lambda^u \J_u \,,
 \end{split} \label{eq:SO55DiffConditions}
\end{equation}
The $\SO{4}_R$ vector $\Lambda_u$ encodes the AdS$_6$ radius and hence cosmological constant as follows. We can use a $\SO{4}_R$ rotation to write, without loss of generality,
\begin{equation}
\Lambda_u = \left( 0, 0, 0, R^{-1} \right) \,,
\end{equation}
with $R$ the AdS$_6$ radius. This breaks the $\SO{4}_R$ to the $\SO{3}_R$ R-symmetry of AdS$_6$ vacua. Let us therefore write $u = \left(A, 4\right)$ with $A = 1, 2, 3$ labelling the vector representation of $\SO{3}_R$. With respect to $\left(A, 4\right)$ the differential conditions become
\begin{equation}
\begin{split}
\gL_{\J_A} \J_B &= - \frac{3}{2^{1/4}\,R}\, \epsilon_{ABC} \J^C \,, \\
\gL_{\J_A} \J_4 &= 0 \,, \\
\gL_{\J_A} \hK &= 0 \,, \\
d\hK &= \frac{2^{3/4}}{R}\, \J_4 \,. \label{eq:SO55RsymmDC}
\end{split}
\end{equation}
Note that the conditions $\gL_{\J_4} \J_u = 0$ and $\gL_{\J_4} \hK = 0$ are automatically satisfied by $\J_4 \propto d\hK$ \cite{Wang:2015hca,Malek:2017njj}.

As we disucssed in section \ref{s:AdSReview}, the $\J_u$ are generalised Killing vector fields since
\begin{equation}
 \gL_{\J_u} \gM_{MN} = 0 \,.
\end{equation}
However, in contrast to the AdS$_7$ case, the equation $\J_4 \propto d\hK$ implies that $\J_4$ is a trivial generalised Killing vector field, containing no spacetime vector field but only exact differential forms. It therefore generates trivial gauge transformations of the gauge potentials. On the other hand, the three generalised vector fields $\J_A$ necessarily contain non-vanishing spacetime vector field components, since $\gL_{\J_A} \neq 0$. These spacetime vector fields must be Killing vector fields that generate the $\SU{2}_R$ algebra and hence are related to an $S^2$ geometry. Therefore, we will consider the internal space
\begin{equation}
 M_{int} = S^2 \times \Sigma \,,
\end{equation}
where $\Sigma$ is a Riemann surface with coordinates $x^\alpha$, $\alpha = 1, 2$. We will, in principle, allow for metrics on $M_{int}$ with off-diagonal components between $S^2$ and $\Sigma$ although we will see that supersymmetry forbids such components.

The $\SL{2}_S$ of the IIB S-duality acts naturally on the Riemann surface and we will raise/lower all $\SL{2}_S$ indices using the alternating symbols $\epsilon_{\alpha\beta} = \epsilon^{\alpha\beta} = \pm 1$ with
\begin{equation}
 \epsilon^{\alpha\gamma} \epsilon_{\beta\gamma} = \delta^\alpha_\beta \,,
\end{equation}
and following a Northwest-Southeast convention. The $S^2$ will once again be parameterised by three functions $y_A$, $A = 1, \ldots, 3$ with $y_A y^A = 1$. Further $S^2$ conventions are described in appendix \ref{A:S2}.

\subsection{$\SO{5,5}$ ExFT and IIB SUGRA}
Supersymmetric AdS$_6$ vacua are characterised by four generalised vector fields $\J_u \in \Gamma\left({\cal R}_1\right)$ and a generalised tensor $\hK \in \Gamma\left({\cal R}_2\right)$. In IIB SUGRA these become formal sums of spacetime vector fields and differential forms as follows
\begin{equation}
 \begin{split}
  \J_u &= V_u + \lambda_u{}^{\alpha} + \sigma_u \,, \\
  \hK &= \homega_{(0)}^\alpha + \homega_{(2)} + \homega_{(4)}^\alpha \,,
 \end{split}
\end{equation}
where $V_u$, $\lambda_u{}^{\alpha}$ and $\sigma_u$ denote the vector, 1-form and 3-form parts of $\J_u$, while $\omega_{(p)}$ are $p$-forms appearing in $\hK$.

The wedge products and tensor products appearing in the algebraic conditions \eqref{eq:AlgConditions} are
\begin{equation}
 \begin{split}
  \J_u \wedge \J_v &= \sqrt{2} \left( \imath_{V_{(u}} \lambda_{v)}^{\alpha} + \lambda_{(u}^{\alpha} \wedge \sigma_{v)} + \left( - \imath_{V_{(u}} \sigma_{v)} - \frac12 \epsilon_{\alpha\beta} \lambda_u^\alpha \wedge \lambda_v^\beta \right) \right) \,, \\
  \hK \otimes \hK \vert_{R_c} &= \homega_{(2)} \wedge \homega_{(2)} + 2\, \epsilon_{\alpha\beta}\, \homega_{(0)}^\alpha\, \homega_{(4)}^\beta \,, \\
  \hK \wedge \K &= \homega_{(2)} \wedge \bomega_{(2)} + \epsilon_{\alpha\beta}\, \homega_{(0)}^\alpha\, \bomega_{(4)}^\beta + \epsilon_{\alpha\beta}\, \bomega_{(0)}^\alpha\, \homega_{(4)}^\beta \,,
 \end{split} \label{eq:IIBAlgConditions}
\end{equation}
where we defined $K = \frac14 \J_u \wedge J^u = \bomega_{(0)}^\alpha + \bomega_{(2)} + \bomega_{(4)}^\alpha$. Moreover, the differential operators appearing in the differential conditions \eqref{eq:SO55DiffConditions} become
\begin{equation}
 \begin{split}
  \gL_{\J_u} \J_v &= L_{V_u} V_v + L_{V_u} \sigma_v + L_{V_u} \lambda_v{}^\alpha \\
  & \quad - \imath_{V_v} d\lambda_u^\alpha - \imath_{V_v} d\sigma_u - \epsilon_{\alpha\beta}\, \lambda_v^\alpha \wedge d\lambda_u^\beta \,, \\
  \gL_{J_u} \hK &= L_{V_u} \homega_{(0)}^\alpha + L_{V_u} \homega_{(2)} + L_{V_u} \homega_{(4)}^\alpha \\
  & \quad + \epsilon_{\alpha\beta}\, \homega^\alpha_{(0)}\, d\lambda_{u}^\beta - \homega_{(0)}^\alpha\, d\sigma_{u} - \homega_{(2)} \wedge d\lambda_u^\alpha \,, \\
  d\hat{K} &= - \sqrt{2}\, d\homega_{(2)} + \sqrt{2}\, d \homega_{(0)}^\alpha \,.
 \end{split} \label{eq:IIBDiffConditions}
\end{equation}

\subsection{Half-maximal structure}
In contrast to AdS$_7$ vacua of mIIA, the gauge potentials of IIB SUGRA on $M_{int} = S^2 \times \Sigma$ can always be chosen to respect the $\SU{2}_R$ symmetry. Therefore, they will automatically be included in the half-maximal structures we construct here.

The most general $\J_A$ we can construct from $\SU{2}_R$-triplets that satisfies the algebraic conditions \eqref{eq:AlgConditions} is, up to generalised diffeomorphisms,
\begin{equation}
 \begin{split}
  \J_A &= \frac{1}{2^{1/4}} \left(\frac{3}{R} \, v_A + 4\,R \left( y_A m^{\alpha} + k^\alpha dy_A \right) + \frac{16\,R^3}{3} \left( |m|\, \theta_A \wedge vol_\Sigma - y_A k^{\beta} m_{\beta} \wedge vol_{S^2} \right) \right) \,,
 \end{split}
\end{equation}
where $v_A$ are Killing vectors and $\theta_A$ certain 1-forms on $S^2$ (see appendix \ref{A:S2}), $m^\alpha = m^\alpha{}_\beta dx^\beta$ are 1-forms on $\Sigma$, where $m^\alpha{}_\beta$ depends only on the coordinates of $\Sigma$, $k^\alpha$ are functions on $\Sigma$, and we defined $|m| = \frac12 m_{\alpha\beta} m^{\alpha\beta}$.

Next, we construct $\hK$ such that is an $\SU{2}_R$-invariant and satisfies $\hK \otimes \hK \vert_{R_c} = 0$ and $J_A \wedge J^A \wedge \hK > 0$. We find the unique combination, up to generalised diffeomorphisms,
\begin{equation}
 \begin{split}
  \hat{K} &= 4\, p_\alpha - \frac{16\,R^2}{3} \left(r + p_\beta k^\beta\right) vol_2 \,, 
 \end{split} \label{eq:SO55KSimple}
\end{equation}
and hence $J_4$
\begin{equation}
 \J_4 = \frac{R}{2^{3/4}} d \hK = \frac{1}{2^{1/4}} \left( 4\,R\, dp^\alpha - \frac{16\,R^3}{3} d \left( r + p_\beta k^\beta \right) \wedge vol_{S^2} \right) \,.
\end{equation}
The algebraic condition
\begin{equation}
 J_A \wedge J^A \wedge \hK = \frac{256\,R^4}{3} r\, |m|\, vol_{S^2} \wedge vol_\Sigma > 0 \,, \label{eq:Ad6Reg}
\end{equation}
is satisfied iff $r\,|m| \geq 0$ with equality at the boundary of $\Sigma$, while the algebraic conditions for $\J_4$
\begin{equation}
 \J_4 \wedge \J_4 = \frac13 \J_A \wedge \J^A \,, \qquad \J_4 \wedge \J_A = 0 \,,
\end{equation}
impose
\begin{equation}
 \begin{split}
  m_{\alpha} \wedge dp^\alpha &= 0 \,, \\
  m^\alpha \wedge m^\beta &= dp^\alpha \wedge dp^\beta \,, \\
  dr + p_\alpha dk^\alpha &= 0 \,.
 \end{split} \label{eq:SO55DiffCond1}
\end{equation}
Note that the final condition can be used to simplify the expression of $\J_4$
\begin{equation}
 \J_4 = \frac{1}{2^{1/4}} \left( 4\,R\, dp^\alpha - \frac{16\,R^3}{3} k_\beta\, dp^\beta \wedge vol_{S^2} \right) \,.
\end{equation}

Finally, we are left to solve the differential conditions \eqref{eq:IIBDiffConditions}. Here these simplify to
\begin{equation}
 \begin{split}
  d\lambda_A^\alpha &= d\sigma_A = 0 \,,
 \end{split}
\end{equation}
which implies $m^\alpha = - dk^\alpha$.

Thus, we find that
\begin{equation}
 \begin{split}
  \J_A &=  \frac{1}{2^{1/4}} \left(\frac{3}{R} v_A + 4\,R\, d\left( k^\alpha\, y_A \right) + \frac{8\,R^3}{3} \rho\, d \left( k^\alpha \theta_A \wedge dk_\alpha \right) \right) \,, \\
  \J_4 &= \frac{1}{2^{1/4}} \left( 4\,R\, dp^\alpha - \frac{16\,R^3}{3}\, k_\beta\, dp^\beta \wedge vol_{S^2} \right) \,, \\
  \hK &= 4\, p_\alpha - \frac{16\,R^2}{3} \left( r + p_\beta k^\beta \right) vol_{S^2} \,, \label{eq:AdS6StructuresSummary}
 \end{split}
\end{equation}
determined entirely by the two $\SL{2}$-doublets of real functions $k^\alpha$ and $p^\alpha$ on $\Sigma$, which satisfy the differential conditions
\begin{equation}
 dk^\alpha \wedge dk^\beta = dp^\alpha \wedge dp^\beta \,, \qquad dk^\alpha \wedge dp_\alpha = 0 \,, \label{eq:AdS6DC}
\end{equation}
and positivity condition \eqref{eq:Ad6Reg}
\begin{equation}
r |dk| \geq 0 \textrm{ with equality at } \partial \Sigma \,, \label{eq:AdS6RegS}
\end{equation}
where $|dk| = \partial_\alpha k_\beta \partial^\alpha k^\beta$, and $r$ is defined up to an integration constant by
\begin{equation}
 dr = - p_\alpha\, dk^\alpha \,. \label{eq:AdS6r}
\end{equation}

At this stage, one might wonder how the quadratic differential conditions \eqref{eq:AdS6DC} can underly supersymmetric AdS vacua, which ought to be described by a first-order BPS equation. The answer is that we still have residual diffeomorphism symmetry on the Riemann surface $\Sigma$ that can be used to turn \eqref{eq:AdS6DC} into first-order differential equations. We will show how to do this after calculating the supergravity fields from the structures.

\subsection{The AdS$_6$ vacua}
We will now compute the supergravity background corresponding to the half-maximal structures \eqref{eq:AdS6StructuresSummary}. The supergravity fields are encoded in the generalised metric \eqref{eq:SO55GenMetric1}, \eqref{eq:SO55GenMetric2} as summarised in appendix \ref{A:IIBfields}. Moreover, the 6-D metric is warped by the factor \cite{Malek:2017njj}
\begin{equation}
 w_6 = |g|^{-1/4} \kappa^2 \,,
\end{equation}
where $|g|$ is the determinant of the internal four-dimensional space. From this, we find the following background in Einstein frame
\begin{equation}
 \begin{split}
  ds^2 &= \frac{\sqrt{2}\,r^{5/4}\,\Delta^{1/4}\,R^2}{3^{3/4} |dk|^{1/2}} \left[ \frac{12}{r} ds_{AdS_6}^2 + \frac{|dk|^2}{\Delta} ds_{S^2}^2 + \frac{4}{r^2} dk^\alpha \otimes dp_\alpha \right] \,, \label{eq:AdS6Vacuum} \\
  C_{(2)}{}^{\alpha} &= - \frac{4\,R^2}{3}\, vol_{S^2} \left( k^\alpha + \frac{r\, p_\gamma\, \partial^\beta k^\gamma\, \partial_\beta p^\alpha}{2\Delta} |dk| \right) \,, \\
  H_{\alpha\beta} &= \frac{1}{2\sqrt{3\,\Delta}} \left( \frac{|dk|}{\sqrt{r}}\, p_\alpha p_\beta + 6 \sqrt{r}\, \partial_\gamma k_\alpha \partial^\gamma p_\beta \right) \,,
 \end{split}
\end{equation}
where
\begin{equation}
 \Delta = \frac34 r\, |dk|^2 + \frac12 |dk|\, p_\gamma p_\delta \partial_\sigma k^\gamma \partial^\sigma p^\delta \,, \qquad |dk| = \partial_\alpha k_\beta \partial^\alpha k^\beta \,,
\end{equation}
and $H_{\alpha\beta}$ is the $\SL{2}$ matrix parameterised by the axio-dilaton $\tau = e^{\psi} + i\, C_0$ as in \eqref{eq:HParam}. The solutions are completely determined by $p^\alpha$ and $k^\alpha$, which are any pair of real $\SL{2}$-doublet functions on $\Sigma$ satisfying \eqref{eq:AdS6DC} and \eqref{eq:AdS6RegS} with $r$ defined by \eqref{eq:AdS6r}.

As we mentioned previously, we can use diffeomorphisms to turn the differential equations for $k^\alpha$ and $p^\alpha$ into first-order PDEs. In particular, we can always use diffeomorphisms to make the metric on $\Sigma$ conformally flat. From \eqref{eq:AdS6Vacuum} we see that this would impose
\begin{equation}
 \partial_1 k^\alpha \partial_1 p_\alpha = \partial_2 k^\alpha \partial_2 p_\alpha \,, \qquad \partial_1 k^\alpha \partial_2 p_\alpha = 0 \,.
\end{equation}
Together with \eqref{eq:AdS6DC}, and requiring \eqref{eq:AdS6RegS} the differential conditions become the Cauchy-Riemann equations
\begin{equation}
 dk^\alpha = I \cdot dp^\alpha \,, \label{eq:CR}
\end{equation}
where $I_\alpha^{\beta} = \delta_{\alpha\gamma} \epsilon^{\gamma\beta}$ is a complex structure on $\Sigma$. This implies that $p^\alpha$ and $k^\alpha$ are the real and imaginary parts of two holomorphic functions on $\Sigma$
\begin{equation}
 f^\alpha = - p^\alpha + i\, k^\alpha \,.
\end{equation}

We now immediately see that our solutions match those of \cite{DHoker:2016ujz} upon identifying our holomorphic functions with the ${\cal A}_{\pm}$ of \cite{DHoker:2016ujz} as follows
\begin{equation}
 {\cal A}_{\pm} = i\, f^1 \pm f^2 \,. \label{eq:HolMatch}
\end{equation}
We give further details of the map between our objects and those of \cite{DHoker:2016ujz} in the appendix \ref{A:AdS6Match}. These solutions can be extended to globally regular solutions by including a boundary of the Riemann surface on which the holomorphic functions $f^\alpha$ have poles as discussed in \cite{DHoker:2017mds,DHoker:2017zwj}.

\section{Minimal consistent truncations} \label{s:ConsTruncation}
As shown in \cite{Malek:2017njj}, we immediately obtain the minimal consistent truncation around the supersymmetric AdS vacua we constructed here. The truncation Ansatz for the scalar fields is given in \eqref{eq:TruncAnsatz}, while that for the remaining fields can be found in \cite{Malek:2017njj}.

\subsection{AdS$_7$} \label{s:AdS7ConsTruncation}
By computing the generalised metric corresponding to the ${\cal J}_u(x,Y)$ and ${\cal \hK}(x,Y)$ in \eqref{eq:TruncAnsatz} we find the truncation Ansatz for the IIA SUGRA fields in string frame
\begin{equation}
\begin{split}
  ds_{10}^2 &= \frac{R^2}{4} \sqrt{-\frac{t}{\ddot{t}}} X^{1/2} ds_7^2 + \frac{R^2}{4} \sqrt{-\frac{\ddot{t}}{t}} \left[ X^{-5/2} d\z^2 + X^{5/2} \frac{t^2}{\dot{t}^2 X^5 - 2\, t\, \ddot{t}} ds_{S^2}^2 \right] \,, \\
  e^{\psi} &= \frac{2}{R} X^{5/4} \left( - \frac{t}{\ddot{t}} \right)^{3/4} \frac{1}{\sqrt{X^5 \dot{t}^2 - 2\, t\, \ddot{t}}} \,, \\
  H_3 &= - \frac{R^2}{4} X^{-5/4} \ddot{t} \left( - \frac{\ddot{t}}{t} \right)^{1/4} \left[ 3 - \frac{t}{\ddot{t}} \frac{m \dot{t}}{\dot{t}^2 X^5 - 2\, t\, \ddot{t}} \right] vol_{\tilde{M_3}} \\
  & \quad - \frac{R^2}{4} X^{-5/4} \ddot{t} \left( - \frac{\ddot{t}}{t} \right)^{1/4} \left(1-X^5\right) \left[ 1 +\frac{4\, t\, \ddot{t}}{\dot{t}^2 X^5 - 2\, t\, \ddot{t}} + \frac{t}{\ddot{t}} \frac{m \dot{t}}{ \dot{t}^2 X^5 - 2\, t\, \ddot{t}} \right] vol_{\tilde{M_3}} \,, \\
  F_2 &= \frac{R^2}{8\sqrt{2}} \left( 2\, \ddot{t} + X^5 \frac{ m\, \dot{t}\, t}{\dot{t}^2 X^5 - 2\, t\, \dot{t}} \right) vol_2 \,,
 \end{split} \label{eq:AdS7CT}
\end{equation}
with the 2-form potential
\begin{equation}
 B_2 = \frac{R^2}{8\sqrt{2}} \left( \z - \frac{\dot{t}\, t\, X^5}{\dot{t}^2 X^5 - 2\, t\, \ddot{t}} \right) vol_2 \,,
\end{equation}
and where $vol_{\tilde{M}_3}$ is the volume form on the internal space of \eqref{eq:AdS7CT}. The truncation Ansatz is completely determined by the functions $t(z)$ satisfying \eqref{eq:tcond}, and corresponds to the truncation Ansatz found in \cite{Passias:2015gya} in the coordinates of \cite{Cremonesi:2015bld}. Upon truncation, $X$ becomes the scalar field of the minimal 7-dimensional gauged SUGRA \cite{Townsend:1983kk}. All of these AdS vacua correspond to the same vacuum of the 7-dimensional theory.

\subsection{AdS$_6$} \label{s:AdS6ConsTruncation}
We can similarly use \eqref{eq:TruncAnsatz} to find the minimal consistent truncation corresponding to the supersymmetric AdS$_6$ vacua of IIB SUGRA we described here and which were previously constructed in \cite{DHoker:2016ujz}. We find in Einstein frame
\begin{equation}
 \begin{split}
  ds^2 &= \frac{\sqrt{2}\,r^{5/4}\,\bDelta^{1/4}\,R^2}{3^{3/4} |dk|^{1/2}} \left[ \frac{12}{r} ds_{AdS_6}^2 + \frac{X^2\,|dk|^2}{\bDelta} ds_{S^2}^2 + \frac{4}{X^2\,r^2} dk^\alpha \otimes dp_\alpha \right] \,, \\
  C_{(2)}{}^{\alpha} &= - \frac{4\,R^2}{3}\, vol_{S^2} \left( k^\alpha + \frac{X^4\,r\, p_\gamma\, \partial^\beta k^\gamma\, \partial_\beta p^\alpha}{2\bDelta} |dk| \right) \,, \\
  H_{\alpha\beta} &= \frac{1}{2\sqrt{3\,\bDelta}} \left( \frac{X^4\,|dk|}{\sqrt{r}}\, p_\alpha p_\beta + 6 \sqrt{r}\, \partial_\gamma k_\alpha \partial^\gamma p_\beta \right) \,,
 \end{split} \label{eq:AdS6Trunc}
\end{equation}
where
\begin{equation}
 \bDelta = \frac34 r\, |dk|^2 + \frac12 X^4\, |dk|\, p_\gamma p_\delta \partial_\sigma k^\gamma \partial^\sigma p^\delta \,,
\end{equation}
and $k^\alpha$, $p^\alpha$ and $r$ satisfy \eqref{eq:AdS6DC}, \eqref{eq:AdS6RegS}, \eqref{eq:AdS6r}. Upon truncation, $X$ becomes the scalar field of the minimal 6-dimensional gauged SUGRA, the so-called $F(4)$ gauged SUGRA \cite{Romans:1985tw}. All these AdS vacua correspond to the same vacuum of the 6-dimensional gauged SUGRA.

\section{Conclusions} \label{s:Conclusions}
In this paper we showed how ExFT can be used to efficiently construct supersymmetric AdS vacua. We focused on supersymmetric AdS$_{7,6}$ vacua of mIIA and IIB SUGRA, respectively, and found the class of infinite solutions desribed in \cite{Apruzzi:2013yva} and \cite{DHoker:2016ujz}. Our method allowed us to immediately derive the minimal consistent truncation around these AdS vacua. We rederived the consistent truncation around AdS$_7$ vacua of mIIA given in \cite{Passias:2015gya}, and found the minimal consistent truncation around the AdS$_6$ vacua of IIB SUGRA.

These consistent truncations are a useful tool in studying the AdS vacua, for example by finding RG flows between different AdS vacua. It would be interesting to explore whether one can keep vector multiplets in the consistent truncation, using the procedure discussed in \cite{Malek:2017njj}, allowing us to differentiate between the various AdS vacua in the lower-dimensional theory.

The method presented here can be generalised to lower dimensions and different amounts of SUSY \cite{Malek:2017njj,Ashmore:2016qvs}, where it may yield new AdS vacua of 10-/11-dimensional SUGRA. Perhaps, it can even be used to provide a classification of supersymmetric AdS vacua of 10-/11-dimensional SUGRA.

Moreover, this formalism is clearly suited to studying the moduli of AdS vacua. For example, \cite{Ashmore:2016oug} showed that in the absence of isometries beyond the $\mathrm{U}(1)_R$, all infinitesimal moduli of ${\cal N}=2$ AdS$_{5}$ vacua can be exponentiated to finite deformations. One could therefore attempt to compute the deformations of the SUGRA backgrounds corresponding to moduli of the AdS vacua. The finite deformations are holographically dual to exactly marginal deformations of the SCFTs and thus by computing the metric on the AdS moduli space, we could holographically determine the Zamolodchikov metric on the conformal manifold.

\section*{Acknowledgements}
The authors thank Alex Arvanitakis, Davide Cassani, Yolanda Lozano, Hagen Triendl, Alessandro Tomasiello, Christoph Uhlemann, Oscar Varela and Daniel Waldram for helpful discussions. EM would also like to thank the organisers of the ``2018 USU Workshop on Strings and Black Holes'' for hospitality while part of this work was completed. EM is supported by the ERC Advanced Grant ``Strings and Gravity" (Grant No. 320045).

\appendix

\section{IIB parameterisation of the $\SO{5,5}$ generalised metric}  \label{A:IIBfields}

Here we give the IIB parameterisation of the $\SO{5,5}$ generalised metric in the $\mbf{16}$ representation. To do this, we decompose $\SO{5,5} \longrightarrow \SL{4} \times \SL{2}_S \times \SL{2}_A$, where $\SL{4}$ is the geometric $\SL{4}$ acting on the internal space, $\SL{2}_S$ generates S-duality and $\SL{2}_A$ is a further $\SL{2}$ group that appears in the decomposition. Accordingly, a generalised vector field in the $\mbf{16}$ decomposes as
\begin{equation}
 V^M = \left( V^{i\,U} ,\, V^{\alpha}{}_i \right) \,,
\end{equation}
where $i = 1, \ldots, 4$, $\alpha = 1, 2$ and $U = +, -$ are fundamental indices of $\SL{4}$, $\SL{2}_S$ and $\SL{2}_A$, respectively.

The generalised metric is given by
\begin{equation}
 \begin{split}
  \mathcal{M}_{+i\,+j} &= e^{1/2} g_{ij} + e^{-3/2} \left( C_{(4)}^2\, g_{ij} + \frac{1}{4} C_{ik\,\alpha}\, \beta^{kr\,\alpha} C_{js\,\gamma} \beta^{st\,\gamma}\, g_{rt} \right) \\
  & \quad -\frac{1}{2} e^{-3/2} C_{(4)} \left( g_{ir}\, C_{jk\,\alpha} \beta^{kr\,\alpha} + (i\leftrightarrow j)\right) + e^{1/2}\, C_{ik\,\alpha}\, C_{jl\,\gamma}\, g^{kl}\, H^{\alpha\gamma} \\
  \mathcal{M}_{+i\,-j} &= e^{-3/2} \left( C_{(4)} \, g_{ij} - \frac{1}{2} C_{ik\,\alpha} \beta^{kl\,\alpha} g_{lj} \right) \,, \\
  \mathcal{M}_{-i\,-j} &= e^{-3/2} g_{ij} \,, \\
  \mathcal{M}_{+i\,\,\alpha}\,{}^j &= e^{-3/2} \left( C_{(4)}\, g_{ik}\, \beta^{jk}{}_\alpha - \frac{1}{2} C_{ik\,\gamma}\, \beta^{kl\,\gamma}\, g_{lm}\, \beta^{jm}{}_\alpha \right) - e^{1/2} C_{ik\,\gamma} g^{kj} H^{\gamma}{}_{\alpha} \,, \\
  \mathcal{M}_{-i\,\,\alpha}\,{}^j &= e^{-3/2} g_{ik}\, \beta^{jk}{}_\alpha \,, \\
  \mathcal{M}_\alpha{}^i{}_\beta{}^j &= e^{1/2} g^{ij} H_{\alpha\beta} + e^{-3/2} \beta^{ik}{}_\alpha\, \beta^{jl}{}_\beta\, g_{kl} \,.
\end{split}
\end{equation}
Here $g_{ij}$ is the 4-d Einstein-frame metric on $M_{int}$, $C_{(4)} = \frac1{4!} \epsilon^{ijkl} C_{ijkl}$ is the dual of the 4-form, $C_{ij\,\alpha}$ denotes the $\SL{2}$-dual of R-R 2-forms and $\beta^{ij}{}_\alpha = \frac12 \epsilon^{ijkl} C_{kl\,\alpha}$ is its dual. Throughout we dualise with $\epsilon^{ijkl} = \pm 1$, the four-dimensional alternating symbol, i.e. the tensor \emph{density}. $H_{\alpha\beta}$ is the $\SL{2}$ matrix parameterised by the axio-dilaton $\tau = e^{\psi} + i\, C_0$,
\begin{equation}
 H_{\alpha\beta} = \frac{1}{\mathrm{Im}\,\tau}
  \begin{pmatrix}
  |\tau|^2 & \mathrm{Re}\,\tau \\ \mathrm{Re}\,\tau & 1
 \end{pmatrix} \,. \label{eq:HParam}
\end{equation}
All our $\SL{2}_S$ indices are raised/lowered by the $\SL{2}$ invariant $\epsilon_{\alpha\beta} = \epsilon^{\alpha\beta} = \pm 1$ in a Northwest/Southeast convention. The $\epsilon_{\alpha\beta}$'s are normalised as
\begin{equation}
 \epsilon_{\alpha\gamma} \epsilon^{\beta\gamma} = \delta_\alpha^\beta \,.
\end{equation}

\section{$S^2$ conventions} \label{A:S2}
We describe the $S^2$ by three functions $y_u$, $u = 1, \ldots, 3$ satisfying
\begin{equation}
 y_u y^u = 1 \,.
\end{equation}
In terms of these functions, the round metric on $S^2$ and its volume form are given by
\begin{equation}
 ds_{S^2}^2 = dy_u dy^u \,, \qquad vol_{S^2} = \frac12 \epsilon_{uvw} y^u\, dy^v \wedge dy^w \,.
\end{equation}
The Killing vectors of the round $S^2$ are given by
\begin{equation}
 v_u^i = g^{ij} \epsilon_{uvw} y^v \partial_j y^w \,,
\end{equation}
where $i, j = 1, 2$ denote a local coordinate basis and $g^{ij}$ is the inverse metric of the round $S^2$. Alternatively, the Killing vectors can be defined as in \cite{Lee:2014mla}.

We also make repeated use of the 1-forms
\begin{equation}
 \theta_u = \epsilon_{uvw} y^v dy^w \,,
\end{equation}
which form a ``dual span'' of the $T^*(S^2)$ to the Killing vectors, i.e.
\begin{equation}
 \imath_{v_u} \theta_v = \delta_{uv} - y_u\, y_v \,.
\end{equation}
Note that the 1-forms $dy_u$, $\theta_u$ and Killing vectors $v_u$ satisfy
\begin{equation}
y_u dy^u = y_u \theta^u = y_u v^u = 0 \,.
\end{equation}

All the objects we introduced above transform naturally under the $\SU{2}_R$ symmetry generated by the Killing vector fields.
\begin{equation}
 L_{v_u} v_v = - \epsilon_{uvw}\, v^w \,, \qquad L_{v_u} y_v = - \epsilon_{uvw}\, y^w \,, \qquad L_{v_u} dy_v = - \epsilon_{uvw} \, dy^w \,, \qquad L_{v_u} \theta_v = - \epsilon_{uvw} \theta^w \,.
\end{equation}

\section{Matching different AdS$_6$ conventions} \label{A:AdS6Match}
Upon imposing the Cauchy-Riemann equations \eqref{eq:CR} and identifying the holomorphic functions as in \eqref{eq:HolMatch}, we find the following match between our objects and those of \cite{DHoker:2016ujz}. To differentiate our $\kappa$ and our parameter $R$ from the objects denoted by the same symbols in \cite{DHoker:2016ujz} we will denote theirs by an underline, $\underline{\kappa}$ and $\underline{R}$. Using \eqref{eq:HolMatch}, we find
\begin{equation}
 \begin{split}
  r &= \frac18 {\cal G} \,, \\
  |dk| &= \frac12 \underline{\kappa}^2 \,, \\
  \Delta &= \frac{3\,{\cal G}\,\underline{\kappa}^4}{128}\left(\frac{1+\underline{R}}{1-\underline{R}}\right)^2 \,, \\
  R^2 &= c \,.
 \end{split}
\end{equation}

To compare our two-forms and axio-dilaton with those of \cite{DHoker:2016ujz}, it is important to translate our $\SL{2}$ representations into their $\SU{1,1}$ ones. These are mapped via
\begin{equation}
 \begin{split}
  {\cal C} &= - C_{2}^1 + i\, C_2^2 \,, \\
  B + \frac{1}{\bar{B}} &= -2 \frac{1 + H_{12}^2 + H_{22}^2}{1 + \left( H_{12} + i\, H_{22} \right)} \,,
 \end{split}
\end{equation}
where ${\cal C}$ denotes the 2-form and $B$ the axio-dilaton of \cite{DHoker:2016ujz}. The latter encodes the complex axio-dilaton $\tau = e^{\psi} + i\,C_0$ as
\begin{equation}
 B = \frac{1+i\,\tau}{1-i\,\tau} \,.
\end{equation}

\bibliographystyle{JHEP}
\bibliography{NewBib}

\end{document}